\documentclass[preprint,showpacs,preprintnumbers,amsmath,amssymb,groupedaddress]{revtex4}

\def\GeVcSq {{\rm (GeV/}c)^2}

\usepackage{graphicx}
\usepackage{epstopdf}
\usepackage{dcolumn}
\usepackage{bm}

\begin{document}

\preprint{}

\title{First Measurement of $A_N$ at $\sqrt{s}=200$~GeV in Polarized Proton-Proton Elastic Scattering at RHIC}
\date{\today}
\author{ S.~B\"{u}ltmann}      
\author{I.~H.~Chiang}
\author{R.~E.~Chrien}
\author{A.~Drees}
\author{R.~L.~Gill}
\author{W.~Guryn}
\author{J.~Landgraf}
\author{T.~A.~Ljubi\v{c}i\v{c}}
\author{D.~Lynn}
\author{C.~Pearson}
\author{P.~Pile}
\author{A.~Rusek}
\author{M.~Sakitt}
\author{S.~Tepikian}
\author{K.~Yip}
\affiliation{Brookhaven National Laboratory, Upton, NY 11973, USA}
\author{J.~Chwastowski} 
\author{B.~Pawlik}
\affiliation{Institute of Nuclear Physics, Cracow, Poland}
\author{M. Haguenauer}
\affiliation{Ecole Polytechnique, 91128 Palaiseau Cedex, France}
\author{A.~A.~Bogdanov}
\author{S.~B.~Nurushev}
\author{M.~F.~Runtzo}
\author{M.~N.~Strikhanov}
\affiliation{Moscow Engineering Physics Institute, Moscow, Russia}
\author{I.~G.~Alekseev}
\author{V.~P.~Kanavets}
\author{L.~I.~Koroleva} 
\author{B.~V.~Morozov} 
\author{D.~N.~Svirida}
\affiliation{Institute for Theoretical and Experimental Physics, Moscow, Russia}
\author{A.~Khodinov}
\author{M.~Rijssenbeek} 
\author{L.~Whitehead}
\author{S.~Yeung}
\affiliation{Stony Brook University, Stony Brook, NY 11794, USA}
\author{K.~De}
\author{N.~Guler}
\author{J.~Li}
\author{N.~\"{O}zt\"{u}rk}
\affiliation{University of Texas at Arlington, Arlington, TX 76019, USA}
\author{A.~Sandacz}
\affiliation{So\l tan Institute for Nuclear Studies, Warsaw, Poland}
\affiliation{}

\collaboration{PP2PP}
\homepage{}

\begin{abstract}
We report on the first measurement of the single spin analyzing power ($A_N$) at $\sqrt{s}=200$ GeV, obtained by the pp2pp experiment using polarized proton beams at the Relativistic Heavy Ion Collider (RHIC).  Data points were measured in the four momentum transfer $t$ range $0.01 \leq |t| \leq 0.03$ $\GeVcSq$. Our result, averaged over the whole $|t|$-interval is about one standard deviation above the calculation, which uses interference between  electromagnetic spin-flip amplitude and hadronic non-flip amplitude, the source of $A_N$. The difference could be explained by an additional contribution of a hadronic spin-flip amplitude to $A_N$.
\end{abstract}
\maketitle
\section{\label{sec:intro}Introduction\protect\\}
The pp2pp experiment~\cite{PP2PPplb04,lynn,guryn} at RHIC is designed 
to systematically study polarized proton-proton ($pp$) elastic scattering 
from \mbox{$\sqrt s =$ 60 GeV to $\sqrt s = $ 500 GeV}, covering 
the $|t|$-range from the region of Coulomb Nuclear Interference (CNI) 
to 1.5~$\GeVcSq$. 
Studies of spin dependence of $pp$ scattering at small momentum transfers and at the highest energies presently available at RHIC offer an opportunity to reveal important information on the nature of exchanged mediators of the interaction, the Pomeron and the hypothetical Odderon (see Ref.~\cite{barone,donnachie} and references therein). The theoretical treatment of small-$t$ scattering is still being developed, hence the experimental data are expected to provide significant constraints 
for various theoretical approaches and models (see Ref. \cite{buttimore} and 
references therein).

In this paper we present the first measurement of the analyzing power 
$A_N$ in $pp$ elastic scattering of polarized protons at RHIC at 
$\sqrt{s} = 200 \:\rm{GeV}$ and $0.01 \leq |t| \leq 0.03$ $\GeVcSq$. $A_N$ 
is defined as the left-right cross section asymmetry with respect to the 
transversely polarized proton beam. In this range of $t$, $A_N$ originates 
mainly from the interference between electromagnetic (Coulomb) spin-flip and 
hadronic (nuclear) nonflip amplitudes \cite{buttimore}. 
However, it was realized that $A_N$ in the 
Coulomb-nuclear interference (CNI) region is a sensitive probe  of the hadronic
spin-flip amplitude \cite{kz}. A possible hadronic single spin-flip 
amplitude would alter $A_N$ and its effect would 
depend on the ratio of the single
spin-flip amplitude ($\phi _5)$ to nonflip amplitudes ($\phi _1$ and $\phi _3$), Eq.(~\ref{eq:r5}):
\begin{equation}
r_5 = m \phi _5 / (\sqrt{-t}\:{\rm{Im}}(\phi _1 + \phi _3) /  2 ),
\label{eq:r5}
\end{equation}
where $m$ is the nucleon mass (see Ref. \cite{buttimore} for definitions).

Other measurements of $A_N$ performed at small $t$ have been obtained at significantly lower energies,  by at least a factor of 10,  than the present experiment. These measurements include recent high precision results from the RHIC polarimeters obtained at $\sqrt{s} = 13.7~\rm{GeV}$  for elastic $pp$ \cite{bravar,jetcal} and 
$pC$ \cite{bravar,carboncal} scattering, as well as earlier results from BNL AGS for 
$pC$ scattering \cite{e950} at $\sqrt{s} = 6.4~\rm{GeV}$ and from FNAL E704 
for $pp$ scattering \cite{e704} at $\sqrt{s} = 19.7~\rm{GeV}$.

The combined analysis of the present result with the earlier ones, especially with the very accurate results of Refs \cite{jetcal,carboncal}, will help to disentangle contributions of various exchange mechanisms involved in elastic scattering in the forward region \cite{trueman}. In particular, such analysis will allow us to extract information on the spin dependence of the diffractive mechanism dominating at high energies.
 
\section{The Experiment}

\begin{figure}
\includegraphics[width=100mm]{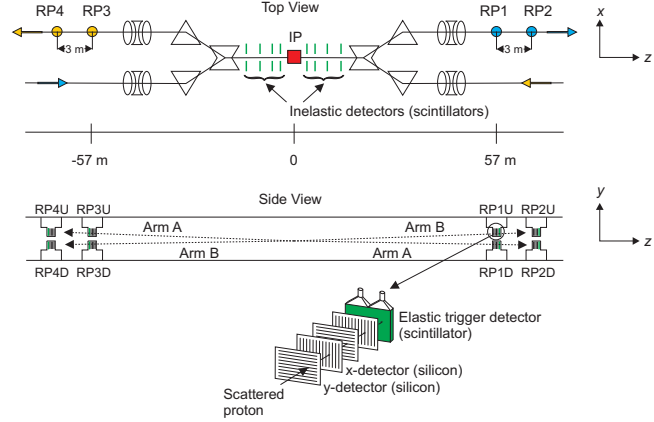}
\caption{\label{layout} Layout of the PP2PP experiment. Note the detector pairs RP1, RP2 and RP3, RP4 lie in different RHIC rings. Scattering is detected in either one of two arms: Arm A is formed from RP3U and RP1D.  Conversely, Arm B is formed from RP3D and RP1U. The coordinate system is also shown.}
\end{figure}

The two protons collide at the interaction point (IP), and since the 
scattering angles are small, scattered protons stay within the beam pipe of 
the accelerator. They follow trajectories determined by the accelerator magnets until they reach the detectors, which measure the $x, y$ coordinates in the plane perpendicular to the beam axis. Those coordinates are measured by Si detectors in the Roman Pots, which are positioned at the location that satisfy so called ``parallel to point focusing". More details on the experiment and the technique used can be found in \cite{PP2PPplb04,lynn}. 
The layout of the experiment is shown in Fig.~\ref{layout}. The identification of elastic events is based on the collinearity criterion, hence it requires the simultaneous detection of the scattered
protons in the pair of Roman Pot (RP) detectors~\cite{battiston}  on either side of the IP.

The elastic event trigger required a coincidence between signals in the RP's scintillators, belonging either to arm A or arm B, see Fig.~\ref{layout}. For each arm the trigger counters in RP1 and RP3 were used. The overall trigger was the logical OR of a coincidence between up (U) and down (D) pots: (RP3U AND RP1D) OR (RP3D AND RP1U) in coincidence with the beam crossing signal derived from the RHIC master clock. %
%
\section{Selection of elastic events}
	      
The detectors in the inner Roman Pots were used for elastic event reconstruction, as this
provided the highest acceptance for the experiment. Particle hits in the silicon detector were identified for each strip  requiring that the energy deposited ($\Delta E$) was $\Delta E \geq 5 \sigma$ of its pedestal value. From those hits a cluster of consecutive strips was formed and the coordinate for that cluster was calculated as an energy-weighted average of the positions of the strips. 



For each RP a hit was formed for an (x,y) coordinate using the clusters in two x planes ($x_1$, $x_2$) and two y planes ($y_1$, $y_2$). A hit required that the distance between two clusters from adjacent planes was $|x_1-x_2| \le 2$ strips, the same for y-coordinate $|y_1-y_2| \le 2$ strips. For matched clusters a single $x$ and $y$ coordinate was calculated as an arithmetic average of the two.  In case there was no match with the second plane one coordinate was used.  

\begin{figure}
\includegraphics[width=100mm]{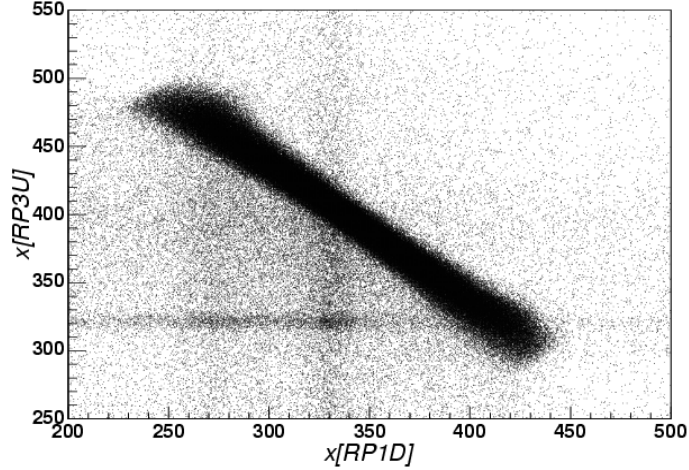}
\caption{Correlation of x-coordinates as measured by the two detectors of arm A before cuts were applied. Note that the background appears enhanced due to the saturation of the main band.}
\label{fig2}
\end{figure}

Because of the collinearity of the scattered protons one has to require a correlation between coordinates measured on each side of the IP. Hence the main criterion to select the elastic scattering events was the hit coordinate correlation in the corresponding silicon detectors on the opposite sides of the  IP.  An example of the correlation of the x-coordinates of the detected protons is shown in Fig.~\ref{fig2}. Note the diagonal band of the elastic events and relatively small background.

Since the events for which the protons were detected in all four RP's allowed reconstructing of the momentum vectors of the scattered protons at the detection point, a subset of those events was used to get better knowledge of the mean coordinates of the collision vertex and of the mean angles of the beams in the IP.  The mean values and widths of those  distributions were also used to determine the correction to the calculated transport matrices, and the beam position at detectors in the horizontal plane. The widths of these distributions are dominated by the beam emittance of about $15\pi$~mm~$\cdot$~mrad for both $x$- and $y$-coordinates and by an uncertainty of about 60~cm (rms) of the vertex position along the beam axis. The latter does not contribute to the width at zero angle scattering in the horizontal plane but contributes significantly at large scattering angles. Thus the $x$-coordinate of  correlation distribution with the minimal width defines the position of tracks scattered at zero angle (or the position of the beam in the detectors) in the horizontal plane. The mean coordinates of non-scattered beams in the detectors were used for planar scattering angle determination instead of the mean coordinates of IP and mean beam angles.  This approach eliminates the contribution of the detector position survey errors, see also discussion of systematic errors.

\begin{figure}
\includegraphics[width=100mm]{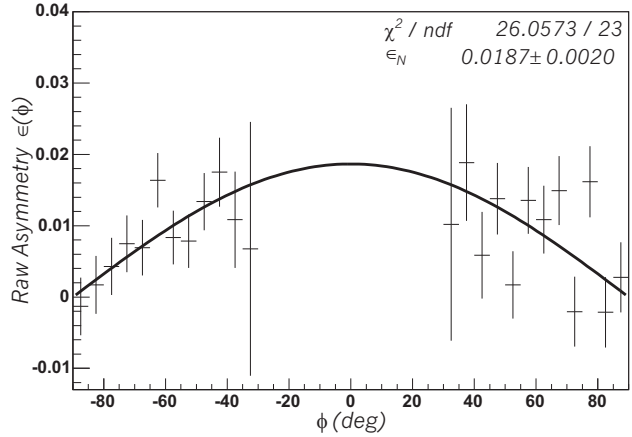}
\caption{The raw asymmetry $\varepsilon(\phi)$ for the full $|t|$-interval.}
\label{fig3}
\end{figure}

To select an elastic event, a match of hit coordinates (x,y) from detectors on the opposite sides of the IP was required to be within $3\sigma$ for x and y-coordinate. The hit coordinates (x,y) of the candidate proton pairs were also required to be in the acceptance area of the detector, determined by the aperture of the focusing quadrupoles located between IP and the RP's. In case that there were more than one match between the hits on opposite  sides of the IP the following algorithm was applied. If there is only one match with number of hits equal to 4, it is considered to be the elastic event. If there is no match with 4 hits or there are more than one such match, the event is rejected.

The average detector efficiency was 0.98, and the upper bound of the elastic events loss  due to all criteria was $\le 3.5\%$. 
 
The background originates from particles from inelastic interactions, beam halo particles and products of beam-gas interactions. The estimated background fraction varies from 0.5\% to 9\% depending on the $y$-coordinate. Since in our analysis the coordinate area was essentially limited to $y > 30$ strips, the background in the final sample does not exceed 2\%. 
			    
\section{Determination of Analyzing Power $A_N$}
			      
After the above cuts, the sample of 1.14 million events, for $N^{\uparrow\uparrow}$  and $N^{\downarrow\downarrow}$ bunch combinations, in the $t$-interval $0.010 \le -t \le 0.030$, subdivided into three intervals $0.010 \le -t < 0.015$,  $0.015 \le -t < 0.020$, $0.020 \le -t \le 0.030$, was used to determine $A_N$.  In each $t$-interval the asymmetry was calculated as a function of azimuthal angle $\phi$ using $5^\circ$-bins. Azimuthal angle dependence of the cross section for the elastic  collision of the vertically polarized protons is given by

\begin{equation}
2\pi\frac{d^2\sigma}{dt d\phi} = \frac{d\sigma}{dt} \cdot
(1 + (P_B + P_Y)A_N\cos{\phi} + P_B P_Y (A_{NN}\cos^2\phi + A_{SS}\sin^2\phi)) \, ,
\label{eq2}
\end{equation}
where $P_B$ and $P_Y$ are the beam polarizations and $A_{NN}$, $A_{SS}$ are double spin asymmetries (see Ref.\cite{buttimore} for definitions). Then the square root formula \cite{ref1} for the single spin raw asymmetry 
$\varepsilon(\phi)$ can be written as
\begin{eqnarray}
\nonumber
\varepsilon (\phi) &=& 
\frac{(P_B + P_Y)A_N\cos{\phi}}{1+P_B P_Y (A_{NN}\cos^2\phi + A_{SS}\sin^2\phi)} \\
&=& 
\frac{\sqrt{N^{\uparrow\uparrow}(\phi)N^{\downarrow\downarrow}(\pi-\phi)}
 - \sqrt{N^{\downarrow\downarrow}(\phi)N^{\uparrow\uparrow}(\pi-\phi)}}
 {\sqrt{N^{\uparrow\uparrow}(\phi)N^{\downarrow\downarrow}(\pi-\phi)} 
 + \sqrt{N^{\downarrow\downarrow}(\phi)N^{\uparrow\uparrow}(\pi-\phi)}} \, .
\label{eq3}
\end{eqnarray}

Beam polarizations for our run were \cite{cad}$P_Y=0.345\pm0.066$ and $P_B=0.532\pm0.106$, leading to an upper constraint of 0.028 for the term $P_B P_Y(A_{NN}\cos^2\phi+A_{SS}\sin^2\phi)$, even if both double-spin asymmetries $A_{NN}$ and $A_{SS}$ were as large as 0.15. This term is small in comparison to the systematic errors on $A_N$ and was therefore neglected in Eq.~(\ref{eq3}) but included in the systematic error, as described below.  A cosine fit to the raw asymmetry $\varepsilon(\phi)$ was used to determine values of $A_N$, see Fig.~\ref{fig3}. 

\section{Systematic Errors}

Equation~(\ref{eq3}), from which the asymmetry is calculated has important features; namely, luminosities of the differently polarized proton beam bunches cancel as do the relative detection efficiencies, including geometrical acceptance, for each $t$ and $\phi$.

However, two other contributions to the systematic error have to be considered: backgrounds, which affect the asymmetry value, and  sensitivity to the transport matrix parameters and to the beam position with respect to the detectors that affect the determination of $t$ and $\phi $.

To check the effect of background, additional selection criteria were applied:
1) rejection of the events with a hit in one of the two y-strips closest to the beam;
2) rejection of events close to the boundary in the ($\phi$,$t$) plane. From these studies, we have found that the upper limit of the systematic error due to the background is $4.5\%$.

The final results were obtained with a transport matrix, which was obtained by correcting the standard transport matrix provided to us by the accelerator physicists from the Collider--Accelerator Department (C--AD). The corrections were calculated
using the fully reconstructed tracks in all four RPs. The results were compared with those  obtained with the standard transport matrix. The relative difference in $A_N$ for the two cases is $1.4\%$. The systematic error due to an uncertainty of beam positions at the detectors is $1.8\%$.

Sensitivity to the variation in $L_{eff}$ was also studied and estimated to be $6.4\%$ assuming upper values of transport uncertainties of $L_{eff}^x$ and $L_{eff}^y$ as large as $\Delta L_{eff}^x/L_{eff}^x =0.1$ and $\Delta L_{eff}^y/L_{eff}^y=0.05$, correspondingly.

As mentioned earlier, neglecting the term with double-spin asymmetries in formula (\ref{eq3}) results in an error  $2.8\%$. 

Since all the above errors are uncorrelated adding them in quadrature results in the systematic error of $\Delta A_N/A_N = 8.4\%$.  This error is smaller than the statistical errors of the measurement, cf. Table~\ref{tab:ANresults}.
 
The polarization values of the proton beams were obtained from the C--AD \cite{cad}. They were evaluated using $A_N$ measurements for elastic proton-Carbon (pC) scattering at small $|t|$-values, in the range $0.01 - 0.02$ (GeV/c)$^2$. The details are described in Ref. \cite{carboncal}. During the period in 2003 when the present data were taken the beam polarizations were $P_Y=0.345\pm0.066$ and $P_B=0.532\pm0.106$.
The errors include the contribution of the systematic part of the error due to the calibration of pC polarimeter of $13\%$, which is correlated for both beams and the uncorrelated statistical errors of the measurement.  This gives the statistical and systematical errors of the measurement in the sum of the polarizations $P_Y+P_B = 0.877 \pm 0.149$.

The total systematic error is comprised of $A_N$ scale error of 17.0$\%$ mostly due to the systematic error of the polarization measurement, and $8.4\%$ error due to the experimental systematic effects as described above. 

An important check of a possible false asymmetry $\varepsilon'$  was obtained from the asymmetry calculated for spin combinations $N^{\uparrow\downarrow}$  and $N^{\downarrow\uparrow}$ with a formula similar to (\ref{eq3}). This term is $\varepsilon'\approx (P'_B - P'_Y ) \cdot A_N$. Given that the polarization values for $N^{\uparrow\downarrow}$  and $N^{\downarrow\uparrow}$ bunches were 
$P'_Y=0.476\pm0.085$ and $P'_B=0.430\pm0.089$ 
 and the $A_N$ in our t-range, one gets $\varepsilon'= -0.0011$ to be compared with the value we measured -0.0016, a good agreement indicating that there is no major source of a false asymmetry. 

\begin{table}[h]
\caption{$A_N$ results} \label{tab:ANresults}
\begin{center}
\begin{tabular}{|l|r|r|r|r|}
\hline
$-t$ interval (GeV/c)$^2$&  0.010--0.015 & 0.015--0.020 & 0.020--0.030 & 0.010--0.030 \\
\hline
$<-t>$ (GeV/c)$^2$ & 0.0127 & 0.0175 & 0.0236 & 0.0185 \\
\hline
$A_N$ & 0.0277 & 0.0250 & 0.0178 & 0.0212\\
\hline
$\Delta A_N$ - stat. & $\pm$0.0061 & $\pm$0.0043 & $\pm$0.0030 & $\pm$0.0023 \\
\hline
$\Delta A_N$ - syst.$^*$ & $\pm$0.0023 & $\pm$0.0021 & $\pm$0.0015&$ \pm$0.0018 \\
\hline
$\Delta A_N$  due to $\Delta(P_Y + P_B)$  & \multicolumn{4}{|c|}{$\pm$17.0 \%}\\
\hline
\multicolumn{5}{|l|}{$^*$ Contributions to systematic error were added in quadrature}\\
\hline
\end{tabular}
\end{center}
\end{table}

\begin{figure}
\includegraphics[width=100mm]{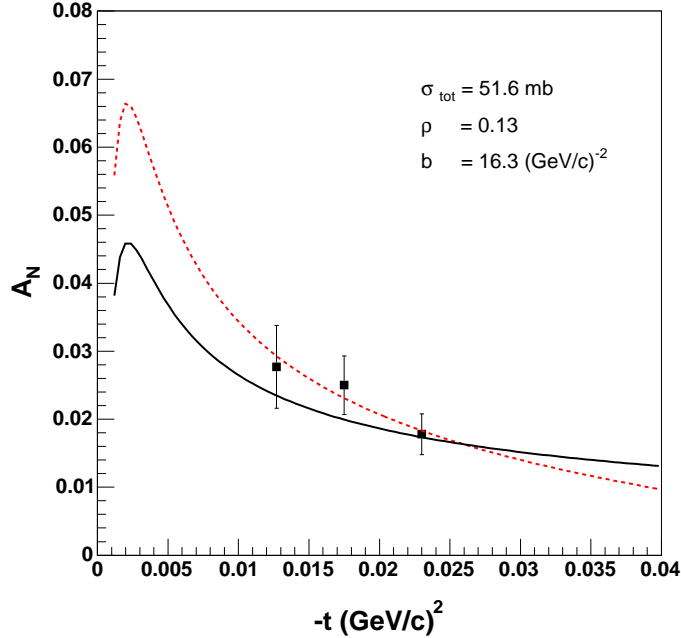}
\caption{The single spin analyzing power $A_N$ for three $t$ intervals. Vertical
error bars show statistical errors. The solid curve corresponds to theoretical calculations without hadronic spin-flip and the dashed one represents the $r_5$ fit.}
\label{fig5}
\end{figure}

\section{Results and Conclusions}
The values of $A_N$ obtained in this experiment and  their statistical errors are shown in  Fig.~\ref{fig5} for the three $t$-intervals, and they are summarized in Table \ref{tab:ANresults}. 

The curves shown in the figure represent theoretical calculations using the 
formula for $A_N$ in the CNI region. The general formula is given by Eq. 28 of
Ref. \cite{buttimore}. With reasonable assumptions that the amplitude $\phi _2$ and the difference
$\phi _1 - \phi _3$ could be neglected at collider energies, the formula 
becomes simpler
\begin{equation}
A_N = \frac {\sqrt{-t}}{m} \: \frac {[\kappa (1 - \rho \:\delta) + 2 (\delta \:{\rm{Re}} \:r_5 - {\rm{Im}} \:r_5)] \frac{t_c}{t} - 2 ({\rm{Re}} \:r_5 - \rho \:{\rm{Im}} \:r_5)}{ (\frac{t_c}{t})^2 - 2 (\rho + \delta)\frac{t_c}{t} + (1 + \rho ^2)} .
\label{cnicurve}
\end{equation}
In this formula $t_c = -8 \pi \alpha /\sigma _{tot}$, $\kappa$ is the anomalous
magnetic moment of the proton, $\rho $ is the ratio of the real to imaginary
parts of forward (nonflip) elastic amplitude, and $\delta $ is the relative
phase between the Coulomb and hadronic amplitudes.
Since the total cross section ($\sigma_{tot}$) and the $\rho$ parameter 
have not been measured in this energy range, we have used values of $\sigma_{tot} = 51.6$ mb and $\rho = 0.13$.  
These values come from fits to the existing $pp$ data taken at energies below 63 GeV and world $p \overline{p}$ data. 
They also agree well with the predictions of various models \cite{block,kopel,compete,bourrely}. The Coulomb phase $\delta$ is calculated  as in Ref. \cite{buttimore},
\begin{equation}
\delta = \alpha\: \ln\:\frac{2}{|t|(b+8/\Lambda ^2)} - \alpha\:\gamma,
\label{delta}
\end{equation}
where $b$ is the slope of the forward peak in elastic scattering, 
$\alpha$ is the fine structure constant, Euler's constant $\gamma = 0.5772$
and $\Lambda ^2 = 0.71 \:\rm{GeV}^2$. 
The value of $b$ comes from our previous measurement \cite{PP2PPplb04}.

\begin{figure}
\includegraphics[width=100mm]{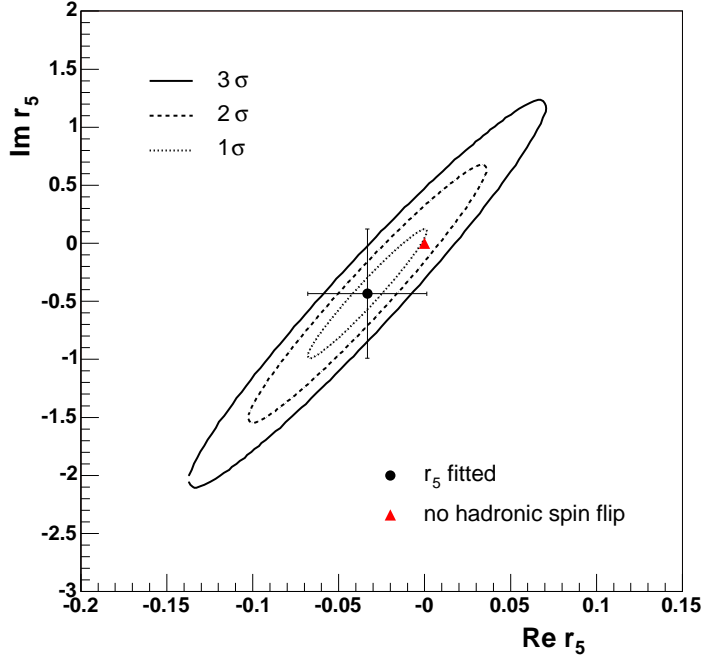}
\caption{Fitted values of $r_5$ (full circle) with contours corresponding to
the different confidence levels. The point corresponding to no hadronic
spin-flip (triangle) is also shown.}
\label{fig6}
\end{figure}

The solid curve in Fig.~\ref{fig5} corresponds to the calculation without hadronic spin-flip (${\rm{Re}} \;r_5$ and ${\rm{Im}} \;r_5$ set to 0 in Eq. \ref{cnicurve}). To quantify a possible contribution of the single helicity-flip amplitude
$\phi _5$, the formula given by Eq. \ref{cnicurve} was fitted 
to the measured $A_N$ values with ${\rm{Re}}\: r_5$ and ${\rm{Im}}\: r_5$ as fit
parameters.
The statistical and systematical errors (except the beam polarization error)
of $A_N$ were added in quadrature for the fit.
The results of the fit are following:  
${\rm{Re}} \:r_5 = -0.033\pm0.035$ and ${\rm{Im}} \:r_5 = -0.43\pm0.56$. The dashed line 
in Fig.~\ref{fig5} respresents the curve resulting from the fit.

The fitted values of ${\rm{Re}} \:r_5$ and
${\rm{Im}} \:r_5$  are shown in Fig.~\ref{fig6} together with contours for 
1$\sigma $, 2$\sigma $ and 3$\sigma $ confidence levels. In addition, the point corresponding to no hadronic spin-flip is also shown. The fitted $r_5$ is compatible, at about one $\sigma $ level, with the hypothesis of no hadronic spin flip. Thus our conclusion is that our results are suggestive of a hadronic spin-flip term, but cannot definitively rule out the hypothesis that only hadronic non spin-flip amplitudes contribute. 

Recent measurements of $A_N$ at substantially lower cms energies than the one reported here indicate small, but significantly different from zero, contribution of spin-flip amplitude in case of proton-carbon scattering~\cite{carboncal,e950} and are consistent with no spin-flip contribution for proton-proton scattering~\cite{jetcal} at $\sqrt{s}$ = 13.7 GeV.



\bibliography{apssamp} 

\begin{thebibliography}{9}
\bibitem{PP2PPplb04} S.~B\"{u}ltmann {\em et al.}, Phys.~Lett.{\bf~B579}, (2004) 245-250.
\bibitem{lynn} S.~B\"{u}ltmann {\em et al.}, Nucl.~Instr.~Meth.~{\bf A535}, (2004) 415-420.
\bibitem{guryn} W.~Guryn {\em et al.}, RHIC Proposal R7 (1994) (unpublished).
\bibitem{barone}
V. Barone, E. Predazzi, {\it High-Energy Particle Diffraction}, Texts and Monographs in Physics, Springer-Verlag; (2002), ISBN: 3540421076.
\bibitem{donnachie}
S. Donnachie, G. Dosch, P. Landshoff, {\it Pomeron Physics and QCD}, Cambridge University Press; (1998), ISBN: B0006Z3XLM.
\bibitem{buttimore} N.~H.~Buttimore {\em et al.}, Phys. Rev. {\bf D59}, 114010 (1999).
\bibitem{kz}
B. Z. Kopeliovich and B.G. Zakharov, Phys. Lett. {\bf B266}, 156 (1989);\\
T. L. Trueman, hep-ph/9610439.
\bibitem{bravar}
    A.~Bravar {\em et al.}, Proceedings of the 16th International Spin
    Physics Symposium {\it SPIN 2004}, eds. F. Bradamante {\it et al}, World Scientific, 2005, p. 700.
\bibitem{jetcal}
    H.~Okada {\em et al.}, Proceedings of the 16th International Spin
    Physics Symposium {\it SPIN 2004}, eds. F. Bradamante {\it et al}, World Scientific, 2005, p. 507.
\bibitem{carboncal}  
O.~Jinnouchi {\em et al.}, Proceedings of the 16th International Spin
    Physics Symposium {\it SPIN 2004}, eds. F. Bradamante {\it et al}, World Scientific, 2005, p. 515.
\bibitem{e950} J.~Tojo {\em et al.}, Phys. Rev. Lett. {\bf 89}, 052302 (2002).
\bibitem{e704}
N.~Akchurin {\it et al}, Phys. Rev. {\bf D48}, 3026 (1993).
\bibitem{trueman}
T.L. Trueman, Proceedings of the 16th International Spin
    Physics Symposium {\it SPIN 2004}, eds. F. Bradamante {\it et al}, World Scientific, 2005, p. 519.
\bibitem{battiston} R.~Battiston {\em et al.}, Nucl.~Instr.~Meth.~{\bf A238}, 35 (1985).
\bibitem{ref1}
   G.G.~Ohlsen and P.W.~Keaton, Jr., Nucl. Instr. Meth. {\bf 109}, 41 (1973).
\bibitem{cad}
O.~Jinnouchi, C--AD Note 171.
\bibitem{block} M. M. Block, Nucl. Phys. {\bf B} (Proc. Suppl.) {\bf 71},
378 (1999).
\bibitem{kopel} B. Z. Kopeliovich, I. K. Potashnikova, B. Povh, 
and E. Predazzi, Phys. Rev. {\bf D63}, 054001 (2001). 
\bibitem{compete} V. V. Ezhela {\em et al.} (COMPETE collab),
Phys. Rev. {\bf D65}, 074024 (2002).
\bibitem{bourrely} C. Bourrely, J. Soffer, and T. T. Wu, 
Eur. Phys. J. {\bf C28}, 97 (2003).
\end{thebibliography}
\begin{acknowledgments}
The research reported here has been performed in part under the US DOE
contract DE-AC02-98CH10886, and was supported by the US National
Science Foundation and the Polish Academy of Sciences. The authors
are grateful for the help of N.~Akchurin, D.~Alburger, P.~Draper,
R.~Fleysher, D.~Morse, Y.~Onel, A.~Penzo, and P.~Schiavon at various stages of the
experiment and for the support of the BNL Physics Department, Instrumentation Division, and the C-AD at the RHIC-AGS facility. We would also like to thank T.~L.~Trueman and B.~Z.~ Kopeliovich for useful discussions.
\end{acknowledgments}

\end{document}